\documentstyle[11pt,newpasp,twoside,epsf]{article}
\def\edcomment#1{\iffalse\marginpar{\raggedright\sl#1\/}\else\relax\fi}
\reversemarginpar

\begin{document}
\title{Warm molecular gas in nearby galaxies. Mapping of the CO(3--2) emission}
\author{Michael Dumke}
\affil{Submillimeter Telescope Observatory (Steward Observatory,
The University of Arizona), 933 N.~Cherry Ave, Tucson, AZ 85721, U.S.A.}

\begin{abstract}
Using the Heinrich-Hertz-Telescope on Mt.\ Graham, Arizona, we have observed
several nearby galaxies in the CO(3--2) line. The CO(3--2) emission
traces -- in contrast to lower rotational transitions -- a warmer and denser
component of the molecular gas. We find that this emission is spatially
extended in the target galaxies.
The excitation conditions seem to vary with the location of the gas within
the galaxy, and also with the overall properties of the target objects.
However, a clear correlation between the CO(3--2) emission and other
quantities, such as Hubble type or IR luminosity, is not present.
\end{abstract}

\section{Introduction}

Since stars form in molecular clouds, the investigation of the molecular
gas component of the ISM is necessary to obtain information about one of
the most important links between the stellar component and the interstellar
medium. The CO(1--0) and (2--1) lines, which are usually used as a tracer
of molecular hydrogen, originate in the large bulk of molecular gas at low
and moderate temperatures, and therefore trace the total amount and
distribution of molecular gas rather than the highly excited molecular
gas close to regions of star formation.

To observe this warmer and denser molecular gas component, one has to
consider higher transitions of CO, like the (3--2) line. Due to the lack
of good sub-mm telescopes at good sites, early observational attempts were
restricted to starburst galaxies, where this line is expected to be strong.
Recently Mauersberger et al.\ (1999) observed a larger sample of galaxies
of various types and activities, but measured only one position for each
object, hence they were not able to account for spatial changes of
excitation conditions {\it within} the galaxies.

In order to obtain a data set which covers several different types of
galaxies and
activity stages {\it and} allows to investigate the gas properties at
different locations within each galaxy, we extensively mapped a total of
twelve nearby galaxies in the CO(3--2) line.

\begin{table*}
\caption{Selected properties of the sample galaxies obtained from
the literature and the CO(3--2) data: Hubble type (col.\ 2,
de Vaucouleurs et al.\ 1991), IR luminosity (col.\ 3, Young et al.\ 1989),
power emitted in the CO(3--2) line (col.\ 4), extent of the central emission
region (col.\ 5), and CO(3--2)/CO(1--0) integrated intensity ratios
(col.\ 6).}
\tabcolsep5pt
\begin{tabular}{llcccc}
\tableline
Source  & Type & $L_{\rm IR}$ & $P_{\rm CO(3-2)}$
	& $l_{\rm x} \times l_{\rm y}$ & $R_{3,1}$ \\
 & & [$\times 10^9\,L_{\sun}$] & [$\times 10^{30}\,{\rm W}$]
	& [kpc $\times$ kpc] & (center/disk) \rule[-4pt]{0pt}{14pt}\\
\tableline
NGC\,253  & SBc    & 15.1 &  5.98 & $0.44 \times 0.16$ & 0.8 / 0.5 \\
NGC\,278  & SBb    &  8.7 &  6.21 &  $2.5 \times 2.4$  & 0.8 \\
NGC\,891  & Sb     & 19.3 &  7.92 &  $1.9 \times 0.3$  & 0.4 / 0.5 \\
Maffei 2  & SBbc   &  --  &  1.08 & $0.37 \times 0.20$ & 1.3 / 0.8 \\
IC\,342   & SBcd   &  2.3 &  0.78 & $0.33 \times 0.19$ & 1.3 / 1.0 \\
NGC\,2146 & SBab\,pec& 85.7 & 61.40 &  $2.4 \times 1.9$ & 1.3 / 1.0 \\
M\,82     & I0     & 29.7 & 16.40 & $0.68 \times 0.34$ & 1.0 / 0.8 \\
NGC\,3628 & SBb\,pec&  6.3 & 13.10 &  $1.4 \times 0.5$  & 1.4 / 1.0 \\
NGC\,4631 & SBd\,sp& 11.9 &  8.33 & $1.75 \times 0.95$ & 1.0 / 0.7 \\
M\,51     & Sbc\,pec& 30.9 & 16.14 &  $2.2 \times 3.3$ & 0.5 / 0.7 \\
M\,83     & SBc    & 15.0 &  7.53 & $0.66 \times 0.55$ & $> 1.2$/ -- \\
NGC\,6946 & SBcd   & 10.8 &  4.21 & $0.48 \times 0.63$ & 1.3 / 1.0 \\
\tableline
\tableline
\end{tabular}
\end{table*}

\section{Observations and data reduction}

The observations were carried out at the
Heinrich-Hertz-Telescope\footnotemark
(Baars \& Martin 1996) on Mt.\ Graham
during several periods between April 1998 and January 2000.
We used a 345\,GHz 2-channel SIS receiver, provided by the MPIfR Bonn,
together with acousto-optical spectrometers. The total bandwidth of the
AOS's is 1\,GHz.
System temperatures during the observations were typically between
600 and 1200\,K.
The data reduction was performed in a standard manner using the
CLASS and GRAPHIC programs of the GILDAS package.
\footnotetext{The HHT is operated by the Submillimeter Telescope Observatory
on behalf of Steward Observatory and the MPI f\"ur Radioastronomie.}
\begin{figure*}
\plotfiddle{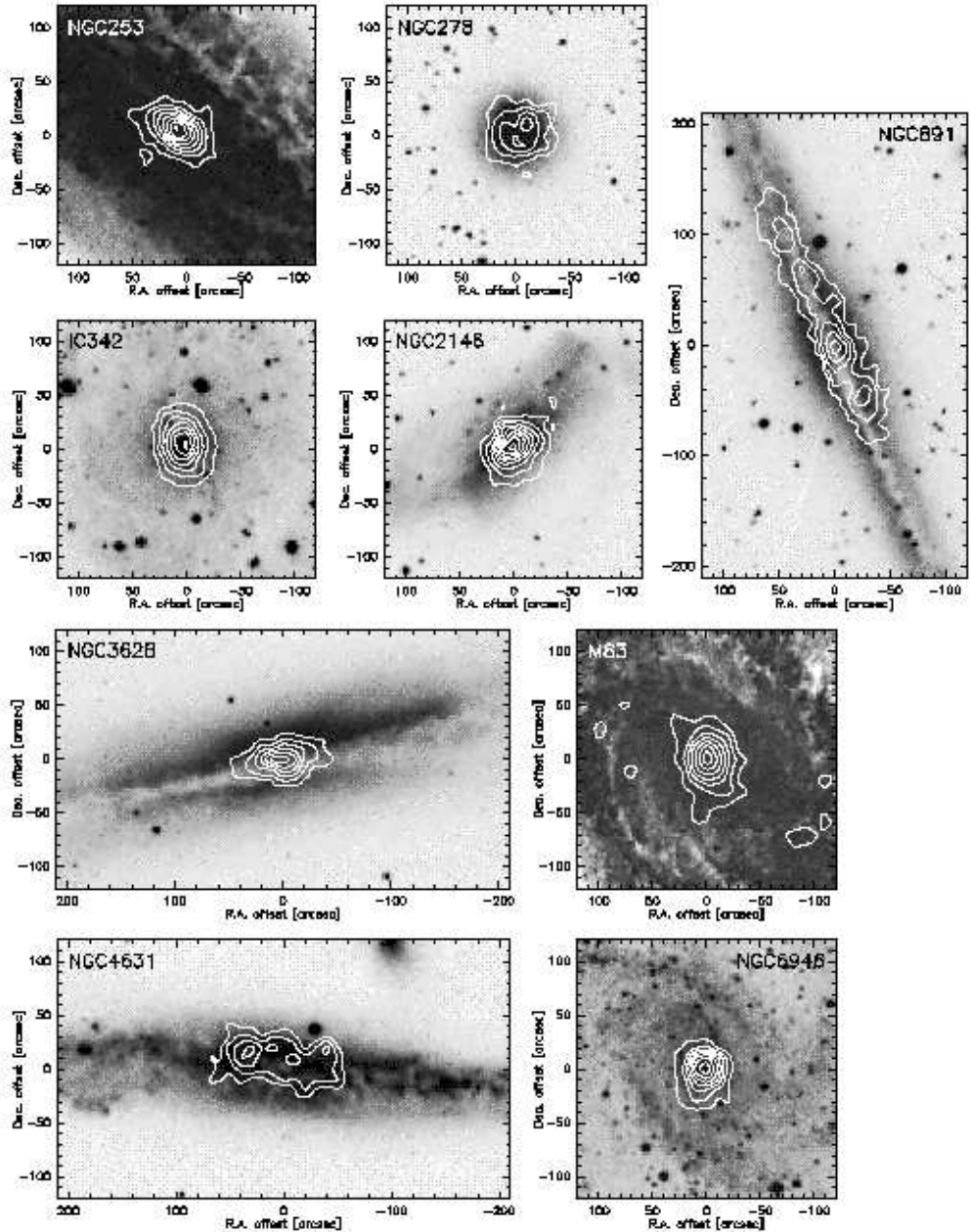}{17.0cm}{0}{74}{74}{-226}{-40}
\caption{Maps of the integrated intensity $I_{\rm CO(3-2)}$ for most of the
galaxies contained in this study, superposed on optical images.
Not shown are maps of M\,51, M\,82, and Maffei\,2 which are available
elsewhere (see text). The contour levels are fractions of the peak
intensity and are therefore different for each object.}
\end{figure*}

\section{Results and discussion}

Of the twelve galaxies which were observed, the results for nine
are shown as integrated intensity maps in Fig.\ 1, overlaid on optical
images extracted from the Digitized Sky Survey. A map of M\,51 is
shown in Wielebinski, Dumke, \& Nieten (1999),
of Maffei\,2 in Walsh et al.\ (in prep.), and CO(3--2) data of M\,82
are available in the literature (e.g.\ Tilanus et al.\ 1991; Wild
et al.\ 1992). In addition a more detailed analysis of the present data,
showing all maps, is given by Dumke et al.\ (2000).

We found that the CO(3--2) emission is not confined to the nucleus
of the galaxies,
but rather extended, with the actual extent depending on the object. In
some cases, the CO(3--2) is as extended as the CO(1--0). In a few
objects, it was even detected in the spiral arms. Nevertheless, the CO(3--2)
emission is more concentrated to the vicinity of star forming structures
(nuclear regions and spiral arms). This is shown by the difference in the
CO(3--2)/(1--0) line ratios between the very centers to regions located
further out. The sizes of the central emission peaks, estimated by
Gaussian fits and deconvolution, are given in Tab.~1.

The CO(3--2) luminosity is enhanced in objects that contain a nuclear
starburst or morphological peculiarities.
The total power emitted in the CO(3--2) line from the central regions
(i.e.\ excluding spiral arms/outer disk) is highest in the starburst
galaxies NGC\,2146, M\,82, NGC\,3628, and in the spiral galaxy M\,51.
When comparing the total power divided by the size of the emission
region, the starbursts M\,82 and NGC\,253 show the highest values
(about $3 - 5$ times higher than the other objects),
and NGC\,278 the smallest.

With the present spatial resolution, the line ratios seem to be independant
of Hubble type, color or luminosity. Most galaxies with enhanced central star
formation show line ratios of $R_{3,1} \sim 1.3$ in the very inner center
and $\sim 1.0$ at a radius of about 1\,kpc. Those objects with a
ring-like molecular gas distribution (M\,82 and NGC\,4631) show lower
ratios. The two galaxies that show CO(3--2) emission distributed
over their spiral arms (NGC\,891 and M\,51) show very low
line ratios despite their high infrared luminosities. This 
result suggests that CO in these two objects reflects a large
amount of molecular gas, rather than enhanced star formation
efficiency.

\section{Outlook}

These observations have shown that it is necessary to study several types
of galaxies in order to obtain results which are not
biased towards starburst objects. Further it is not
sufficient to observe only one point per galaxy, since this cannot account
for differences of the CO(3--2) spatial distribution.

In order to improve this study, our group will survey
more galaxies in the future. This will lead to better statistics
of the findings presented here. Even higher transitions of
$^{12}$CO are considered, as well as observations of other isotopomers
to restrict interpretations concerning optical depth effects.

\acknowledgements
I thank my colleagues R. Beck, Ch.\ Nieten, G. Thuma, R. Wielebinski,
and W. Walsh
for the collaboration during the observations. Further I thank the
SMTO staff for their help at the telescope.

\end{document}